\documentclass[aps,twocolumn,showpacs,groupedaddress,amsmath]{revtex4}

\usepackage[dvips]{graphicx}
\usepackage{dcolumn}
\usepackage{longtable}
\usepackage{bm}

\begin{document}
%\preprint{}

%\widetext

\title{ Geminal wavefunctions with Jastrow correlation: a 
first application to atoms} 

\author { Michele Casula and Sandro Sorella}

\affiliation{International School for Advanced Studies (SISSA)
Via Beirut 2-4, 34014 Trieste, Italy  \\ and INFM Democritos National Simulation
Center, Trieste, Italy}

\date{\today}

\begin{abstract}
We introduce a simple generalization of the well known geminal wavefunction 
already applied in Quantum Chemistry to atoms and small molecules. 
The main feature of the proposed wavefunction is the presence of 
the antisymmetric geminal part together with a Jastrow factor.
Both the geminal and the Jastrow
play a crucial role in determining the remarkable accuracy of the many--body
state: the former permits the correct treatment of the nondynamic correlation
effects, the latter allows the wavefunction to fulfill the cusp conditions and
makes the geminal expansion rapidly converging to the lowest possible
variational energies. This ansatz is expected to provide a substantial part 
of the correlation energy for general complex atomic and molecular systems.
The antisymmetric geminal term can be written as a
single determinant even in the polarized cases.
In general, therefore,  the computational
effort to sample this  correlated wavefunction  
is not very demanding, the scaling of the algorithm with the number 
of atoms  being comparable with  
the simplest Hartree-Fock calculation. 

We applied this Jastrow-geminal approach 
  to atoms up to $Z=15$, always getting 
good variational energies,  by  particularly  improving  
those with a strong multiconfigurational nature.
Our wavefunction is very useful for Monte Carlo techniques, such 
as Fixed node. Indeed,  the nodal surface obtained within  this 
approach can be substantially  improved through the geminal expansion.
\end{abstract}


\maketitle

\section{introduction}

One of the main goals in electronic structure calculations is to deal with a
wavefunction both accurate to predict the physical properties of a quantum system
and simple enough to allow feasible computations of them.
In particular, although
multideterminantal CI like methods could be in principle very accurate, 
in practice the determinantal expansion becomes
heavier and heavier from the computational point of view,
as the number of determinants dramatically increases with the complexity of
the electron system. On the other hand, a single
determinantal wavefunction, kernel of methods like Hartree--Fock (HF) and
Density Functional Theory (DFT), is sometimes
not sufficient to describe 
strongly correlated systems, 
as for instance the transition metal compounds
and the near degenerate shell structure of some atoms.

Since 50's, the intensive efforts to explain theoretically the
superconductivity have been highlighting the role of pairing in the electronic
structure. The BCS wavefunction belongs to an original ansatz in which the
correlation is introduced through the product of pairing functions (in this
context called Cooper pairs), already exploited in quantum chemistry by the
pioneering work of Hurley \textit{et al.} \cite{hurley} to treat correlation effects
in molecular properties. Their wavefunction was called \emph{antisymmetrized
geminal power} (AGP) that has been shown to be the particle--conserving
version of the BCS ansatz \cite{schrieffer}. It includes the single
determinantal wavefunction, i.e. the uncorrelated state, as a special case
and introduces correlation effects in a straightforward way, through the
expansion of the pairing function (geminal): therefore it was studied as a
possible alternative to the other multideterminantal approaches. Although this
ansatz seemed so appealing, it led to some expensive optimization procedures
\cite{bratoz,weiner} with numerical problems \cite{bessis,kurtz} 
in particular when applied to large systems, 
and so it turned out to be non competitive with respect to HF and CI. 

We show in this paper that the use of 
Monte Carlo methods can overcome the previous difficulties in optimizing the 
AGP wavefunctions. 
Two of the most appealing features of these techniques are the possibility
to tackle in a smart way the many--body interacting problem,
having the freedom in the choice of the functional form of the wavefunction,
and to implement very efficient projection algorithms, like diffusion Monte Carlo (DMC) and
Green's function Monte Carlo (GFMC). The trial wavefunction used in these
methods is obtained by multiplying an antisymmetric term
(usually called Slater term) to a Jastrow factor, which correlates the
electrons and takes into account the interelectron
cusp conditions the true wavefunction must fulfill. The Slater term can be
either HF or CI or AGP like. As already pointed out by Umrigar
\cite{umrigar}, the rate of convergence of CI expansion is increased by the
Jastrow factor, just because it allows the wavefunction to have the correct
cusps, otherwise present only asymptotically in the linear combination of
determinants.

In this paper we study the AGP--Jastrow ($AGP+J$) wavefunction, applied to the atomic
problem. The wavefunction we consider is actually a Resonating Valence Bond (RVB)
state, first investigated in 1973 by Anderson \cite{anderson,anderson2}
to study the
high--temperature superconductivity and later applied by Bouchaud and
Lhuillier \cite{bouchaud0} in Monte Carlo calculations of liquid $^3He$.
It is remarkable that this kind of wavefunction can describe well the ground
state of all the atoms studied here, in particular those light atoms affected
by nondynamic correlation.
Moreover, the Jastrow part is
crucial not only in reducing the pairing
expansion but also in inducing a significant improvement of the nodal 
surfaces of the wavefunction, 
once we optimize both the Jastrow and the determinantal part at the
same time.
The minimization of the total energy can be efficiently performed by using
an optimization procedure based upon the Stochastic Reconfiguration (SR) 
method \cite{sorella}, which allows us to determine a large number of
variational parameters, both for the Jastrow and the determinantal part.
Due to the simplicity of the $AGP+J$ wavefunction and its
capability to take into account resonating Slater configurations,
this wavefunction is expected to be appropriate  
also for molecules and more complex systems. 

The paper is organized as follows: in Sec. \ref{form} we define the
wavefunction, in Sec. \ref{method} we describe the energy minimization
method and finally in Sec. \ref{results} and \ref{conclusions} we present
detailed results and conclusions.

\section{functional form of the wave function}

\label{form}

The wavefunction we used in our QMC calculations is the antisymmetrized
product of geminals ($AGP$), multiplied by a simple symmetric Jastrow factor:
\begin{equation}
\Psi(\mathbf{r}_1,\dots,\mathbf{r}_N)=\Psi_{AGP}(\mathbf{r}_1,\dots,\mathbf{r}_N)
J(\mathbf{r}_1,\dots,\mathbf{r}_N) ,
\label{wave}
\end{equation}
where the origin of the reference frame is the nuclear position.
The two parts of the wavefunction (\ref{wave}) will be described in detail
below.

\subsection{The determinantal part} 

Let $\Phi$ be the pairing function (geminal) which takes into account the
correlation between two electrons with opposite spin. If the system is
unpolarized and the state is a spin singlet, the $AGP$ wavefunction is
\begin{equation}
\mathbf{\Psi}_{AGP}(\mathbf{r}_1,\dots,\mathbf{r}_N)= \hat{\textit{A}}
[\Phi(\mathbf{r}^\uparrow_1,\mathbf{r}^\downarrow_2)
\Phi(\mathbf{r}^\uparrow_3,\mathbf{r}^\downarrow_4) \cdots
\Phi(\mathbf{r}^\uparrow_{N-1},\mathbf{r}^\downarrow_N)] , 
\end{equation}
where $\hat{\textit{A}}$ is an operator that antisymmetrizes the product in
the square brackets and the geminal is a singlet:
\begin{equation}
\Phi(\mathbf{r}^\uparrow,\mathbf{r}^\downarrow)=\phi(\mathbf{r}^\uparrow,\mathbf{r}^\downarrow)
\frac{1}{\sqrt{2}} \left ( |\uparrow \downarrow \rangle - |\downarrow \uparrow
\rangle \right ),
\end{equation}
implying that $\phi(\mathbf{r},\mathbf{r}^\prime)$ is symmetric under a
permutation of its variables.
Given this conditions, one can prove \cite{bouchaud} that the spatial part of
the $\mathbf{\Psi}_{AGP}$
can be written in a very compact form:
\begin{equation}
\Psi_{AGP}(\mathbf{r}_1,\dots,\mathbf{r}_N)=det(A_{ij}),
\label{compact}
\end{equation}
where $A_{ij}$ is a $\frac{N}{2} \times \frac{N}{2}$ matrix defined as:
\begin{equation}
A_{ij}=\phi(\mathbf{r}^\uparrow_i,\mathbf{r}^\downarrow_j).
\end{equation}

We are going to extend the definition of the geminal wavefunction to a
polarized system, i.e. a system with a different number of electrons for each
spin. This generalization of the geminal model
was first proposed by Coleman \cite{coleman} and called
$GAGP$.
Without loss of generality,
one can assume that the number of up particles ($N^\uparrow$) is greater than
the down ones ($N^\downarrow$). In order to write down the many--body wavefunction 
with the geminals, one needs to introduce $N^\uparrow-N^\downarrow$
single particle spin orbitals $\overline{\Phi}_j$ not associated with any
geminal and holding unpaired electrons.
Once again one recovers the compact notation (\ref{compact}) for the spatial
part of $\Psi_{AGP}$ (see Appendix), but this time $A_{ij}$ is a
$N^\uparrow \times N^\uparrow$ 
matrix defined in the following way:
\begin{equation} \label{finaldet}
A_{ij} = \left \{
%{ 
\begin{array}{rl}
\phi(\mathbf{r}^\uparrow_i,\mathbf{r}^\downarrow_j) & \textrm{~~~for
$j=1,N^\downarrow$ } \\
%\atop
\overline{\phi}_j(\mathbf{r}^\uparrow_i) & \textrm{~~~for
$j=N^\downarrow+1,N^\uparrow$} \\
\end{array}
%}
\right . 
\end{equation}
where the index $i$ ranges from $1$ to $N^\uparrow$. 

The pairing function can be expanded over a basis of single particle orbitals
\cite{lowdin}:
\begin{equation} \label{geminal}
\phi(\mathbf{r}^\uparrow,\mathbf{r}^\downarrow)=\sum_{i=1}^M \lambda_i
\psi_i(\mathbf{r}^\uparrow) \psi_i^*(\mathbf{r}^\downarrow),
\end{equation}
where $\psi_i$ are general real or complex 
normalized  functions and $\lambda_i$ 
are variational parameters that may depend on the chosen  spatial
symmetry of the geminal. 
Hereafter, for simplicity, 
we do not assume that the orbitals are mutually orthogonal. 

For the application to atoms, in order the wavefunction $\Psi$ has a definite 
angular momentum, it is convenient that the  
geminal is rotationally invariant around the nucleus. This 
requirement  is achieved by taking the generic 
orbital $\psi_i$  to be an eigenfunction of the single particle angular
momentum operators $l^2$ and $l_z$; hence, the orbital will be denoted by:
\begin{equation}
\psi_{nlm}(r,\theta,\phi)=R_{nl}(r) Y_{lm}(\theta,\phi),
\end{equation}
where $Y_{lm}$ are spherical harmonics with standard notations and the 
radial part $R_{nl}$ depends on the principal quantum number $n$.
In this way, the atomic geminal function takes the form:
\begin{equation} \label{geminal2}
\phi(\mathbf{r}^\uparrow,\mathbf{r}^\downarrow)=\sum_{nl}
\sum_{m=-l}^l \lambda_{nl} (-1)^m
\psi_{nlm}(\mathbf{r}^\uparrow) \psi_{nlm}^*(\mathbf{r}^\downarrow).
\end{equation}
In the polarized case, the remaining orbitals $\overline \psi_j$ 
 may change  the 
total angular momentum and spin  quantum numbers 
with the same rules valid for Slater-type wavefunctions. 
    Within our ansatz it is therefore possible to have  
definite total spin and angular momenta at least in all cases when the conventional 
 Slater function does. 

The minimal  geminal expansion is for
$M=N^\downarrow$, then the $AGP$ wavefunction is reduced to the HF
one; instead, if $M$ is greater than $N^\downarrow$, one can prove
that the $AGP$ is equivalent to a linear combination of many Slater
determinants. Therefore, within this functional form, one is able to introduce
the correlation in the determinantal part of the wavefunction just by adding
some terms in the geminal expansion. For a two--electrons
closed--shell system, like Helium atom, or an ensemble of such noninteracting
systems, this wavefunction is exact. For the other atoms, we will show that the
$AGP+J$ wavefunction can lead to very good variational energies,
even in the cases where the HF approximation is rather poor, especially for
light elements.

In order to optimize the radial part $R_{nl}$ of the 
the single particle orbitals, we expand these radial functions in 
a  Slater basis, in close  analogy with Roothaan--Hartree--Fock calculations \cite{roetti},
namely using functions of the type: 
\begin{equation}
r^{n-1} e^{-z_k r}
\end{equation}
with $n\ge 1$, taking in principle as many different $z_k$'s 
as required for convergence. 

In the Roothaan--Hartree--Fock
the coefficients of the linear combinations are more involved, since 
the orthogonality among all single particle states is 
required.
In the Monte Carlo approach we have found that it is much simpler and 
more efficient to deal with non-orthogonal orbitals,  
without spoiling the accuracy of the calculation. 
In fact, for light elements with $Z \le 15$, studied here, it is possible to obtain
almost converged results  by using only two 
exponentials for each radial component (double zeta).

 Hence, our single particle
orbitals read in general
\begin{equation}
R_{nl}(r)=C r^{n-1} ( e^{-z_1 r } + p e^{-z_2 r } ),
\label{orb}
\end{equation}
where $p$ is another variational parameter and $C$ is the normalization
factor for  the radial part $R_{nl}$:
\begin{widetext}
\begin{equation}
C = \frac{1}{\sqrt{(2n)!}} \left( {(2z_1)}^{-(2n+1)} + 2p{(z_1+z_2)}^{-(2n+1)} +
p^2{(2z_2)}^{-(2n+1)} \right)^{-\frac{1}{2}}.
\end{equation}
\end{widetext}
Actually $p$ is not free for all the orbitals: indeed, 
for a more accurate variational wavefunction, we impose the
electron--nucleus cusp condition \cite{kato,pack}, which is  
 satisfied by the exact many body ground state and implies that 
each orbital must fulfill the following relation:
\begin{equation}
\frac{\partial \hat{\psi}}{\partial r} = - Z \psi 
\label{cusp}
\end{equation}  
at $r=0$ (the hat denotes the spherical average). That condition is 
automatically obeyed by all but $1s$ and $2s$ orbitals of the type
given in Eq.\ref{orb}. Instead, the parameter $p$ of $1s$ orbital must
be:
\begin{equation}
p=\frac{z_1-Z}{Z-z_2},
\end{equation}
and for the $2s$ state, 
in order to fulfill Eq. (\ref{cusp}), we choose a
functional form of the type:
\begin{equation}
\psi_{2s}(r)=e^{-z_1 r} + (p + \alpha r) e^{-z_2 r},
\label{cusp2}
\end{equation}
where $\alpha$ is a  further 
variational parameter and $p$ is  given by
\begin{equation}
p=\frac{z_1-\alpha-Z}{Z-z_2}.
\end{equation}
In our study,  we found that the presence of the $\alpha$ term leads to a very
marginal improvement of the variational wavefunction, therefore we set
$\alpha=0$ and we kept the $1s$ and $2s$ orbitals to have the same functional
form, in order to reduce at most the total number of parameters. 

\subsection{The Jastrow factor}

The Jastrow factor in our wave function is very simple and has been widely
used in previous Monte Carlo electronic structure calculations:
\begin{equation}\label{jastrow}
J(\mathbf{r}_1,\dots,\mathbf{r}_N) = \prod_{i < j}  \exp \left( f(\mathbf{
r}_i,\mathbf{r}_j) \right ) 
\end{equation}
where the product is over all pairs of electrons
and  for simplicity  the function  $f$ depends  only 
on their relative  distance $r_{ij}$ and their spins
 $\sigma_i$ and $\sigma_j$, namely: 
\begin{equation} \label{jastrows} 
f_{two~body}(r_{ij})=\frac{a_{\sigma_i \sigma_j} 
r_{ij}}{1 + b _{\sigma_i \sigma_j} r_{ij}} .
\end{equation}
The value of $a_{\sigma_i \sigma_j}$ is fixed by
the cusp condition at the coalescence points of two electrons 
and $b_{\sigma_i \sigma_j}$ contains at 
most three free variational
parameters, as $b_{\uparrow \downarrow}=b_{\downarrow \uparrow}$  
is implied by the spatial symmetry of the Jastrow factor.
The cusp conditions for parallel and antiparallel spin electrons are different
and  yield two different values  of $a_{\sigma_i \sigma_j}$:
\begin{subequations} 
\begin{equation}
a = \frac{1}{2} \text{~for~} \sigma_i=\sigma_j
\label{antiparallel}
\end{equation}
\begin{equation}
a = \frac{1}{4} \text{~for~} \sigma_i \ne \sigma_j.
\label{parallel}
\end{equation}
\end{subequations}
 As pointed out in   Ref.~\onlinecite{filippi&umrigar},
whenever   $a_{\sigma_i \sigma_j}$  or $b_{\sigma_i \sigma_j}$ 
depend on the electron spins  $\sigma_i ~ \sigma_j$,
the wavefunction will  be \emph{spin contaminated}, i.e. it will 
 not be an eigenstate of the
total spin operator $S^2$.
 We have chosen  to preserve the correct spin symmetry of
the total wavefunction, by keeping $a_{\sigma_i \sigma_j}=1/2$ and 
$b_{\sigma_i \sigma_j}=b$, hence fulfilling only the first 
condition (\ref{antiparallel}). 
Indeed, the cusp condition for electrons with parallel spins 
is much less important because  
 their  probability to get   close
 is clearly small, due to the Pauli principle. 

For the nitrogen atom, 
we checked the quality of this wavefunction with respect to a spin contaminated 
one with two variational parameters,
$b_{\uparrow \uparrow}=b_{\downarrow \downarrow}=b_1$ 
and $b_{\uparrow \downarrow}=b_2$, both of them reported in Table \ref{slater}, 
listed with the other atoms. 
In both the cases, we kept the geminal expansion to be minimal ($HF+J$ like
wavefunction). 
 As reported in 
Table \ref{energy},
the improvement in energy obtained by contaminating the spin wavefunction 
is rather negligible, and disappears when the FN DMC simulation is carried out.
This implies that it is possible to obtain almost optimal nodes, without 
spoiling the spin symmetry and by using only one variational parameter for the 
Jastrow factor. 

In order to study the effectiveness of the 
Jastrow for a more accurate determination of the nodal surface,
we have implemented a more involved  Jastrow factor, 
including three-body correlation terms, and we have applied it to few atoms ($Be$
and $Mg$).
Indeed the Jastrow term  $f$ in Eq. \ref{jastrow} is a general function of 
the positions of two  electrons and it has been parametrized 
similarly to the geminal function (\ref{geminal}), but
truncated up to the $l=1$ angular momentum:
\begin{eqnarray}
f(\mathbf{r}_i,\mathbf{r}_j) & = & f_{two~body}(r_{ij}) + \nonumber \\  
& & \psi_0(\mathbf{r}_i) \psi_0(\mathbf{r}_j) + 
\overline{\psi}_1(\mathbf{r}_i) \cdot \overline{\psi}_1(\mathbf{r}_j) .
\end{eqnarray}
The functional form for the s-wave ($\psi_0$) and p-wave ($\overline{\psi}_1$) components
can be chosen among different types; the most widely used in the literature
are the polynomial \cite{wilson,sarsa} and the gaussian--polynomial 
mixed form \cite{savin}. In this work, we have selected the expansion over a
gaussian basis, as reported in Table \ref{agpslater3}.
 This parametrization of the Jastrow factor, 
though being certainly less general compared with  the 
 best possible ones \cite{huang},  includes the most significant 
part of the three-body correlation \cite{smidth}, which involves two electrons and the nuclei.
Our purpose,  in fact, is to check whether it is possible to
lower significantly the energy of the  
 $AGP+J$  wavefunction,  whenever  the Jastrow part of the wavefunction 
is systematically improved together with the determinantal part.

\section{method} 
\label{method}
\subsection{Minimization method}
We have performed the wavefunction optimization by using the
\emph{stochastic minimization} of the total energy 
based upon the Stochastic Reconfiguration (SR)
technique, already exploited for lattice systems \cite{sorella}.
Let $\Psi_T(\alpha^0)$ be the wavefunction depending on an initial set of
$p$ variational parameters $\{\alpha^0_k\}_{k=1,\ldots,p}$.
 Consider now  
 a small variation of the parameters $\alpha_k = \alpha^0_k + \delta \alpha_k$.
The  corresponding  wavefunction $\Psi_T (\alpha)$ 
is equal, within the validity of the linear expansion, to the following one: 
\begin{equation}
\Psi_T^\prime(\alpha) =   \Big(\Psi_T (\alpha^0)  + \sum_{k=1}^p 
\delta\alpha_k ~\frac{\partial}{\partial \alpha_k}\Psi_T (\alpha^0) \Big)
\end{equation}
 Therefore, by introducing local operators  defined on each 
configuration $x=\{ \mathbf{r}_1, \ldots , \mathbf{r}_N \}$ 
as the logarithmic derivatives with respect to the variational
parameters:
\begin{equation}
O^k(x)= \frac{ \partial }{ \partial \alpha_k } \ln \Psi_T (x) 
\end{equation}
and for convenience the identity operator $O^0=1$, we can 
write $\Psi_T^\prime$ in a more compact form:
\begin{equation}\label{subspace}
| \Psi_T^\prime(\alpha) \rangle =  \sum_{k=0}^p 
\delta\alpha_k  O^k | \Psi_T \rangle ,
\end{equation}
where $| \Psi_T \rangle = | \Psi_T ( \alpha_0) \rangle$ and
$\delta\alpha_0=1$. In general, $\delta \alpha_0 \neq 1$, in that case the variation
of the parameters will be scaled
\begin{equation}
\delta \alpha_k \rightarrow \frac{\delta\alpha_k}{\delta \alpha_0}
\end{equation}
and $\Psi_T^\prime $ will be proportional to $\Psi_T(\alpha)$ for small
$\frac{\delta \alpha_k}{\delta \alpha_0}$.
  
 Our purpose is to set up an iterative scheme to reach the minimum
possible energy for the parameters $\alpha$, exploiting the linear
approximation for $\Psi_T(\alpha)$, which will become more and more accurate
close to the convergence, when the variation of the parameters is smaller and
smaller. We follow the stochastic reconfiguration method and define 
\begin{equation} \label{green} 
|\Psi_T^\prime \rangle = P_{SR} (\Lambda  -H)  |\Psi_T \rangle 
\end{equation}
where $\Lambda$ is a suitable large shift, allowing $\Psi_T^\prime$ 
to have a lower energy than $\Psi_T$ \cite{sorella}, and $P_{SR}$ is a projection 
operator over the ($p+1$)--dimensional subspace, spanned by the basis $\{ O_k |
\Psi_T \rangle \}_{k=0,\ldots,p}$, over which the function $|\Psi_T^\prime
\rangle$ has been expanded (Eq. \ref{subspace}).  
Indeed, in order to determine the coefficients $\{\delta\alpha_k
\}_{k=1,\ldots,p}$ corresponding to $\Psi_T^\prime$ defined in Eq.\ref{green},
one needs to solve the SR conditions:
\begin{equation} \label{srconditions}
\langle \Psi_T|O^k (\Lambda -H) |\Psi_T \rangle=
\langle \Psi_T|O^k|\Psi_T^\prime \rangle \textrm{~~for~} k=0, \ldots,p
\end{equation}
that can be rewritten in a linear system:
\begin{equation}
\sum_l \delta\alpha_l~ s^{l k} = f^k,
\label{sys}
\end{equation}
where $s^{l k} = \langle \Psi_T|O^l O^k| \Psi_T \rangle$ is the covariance matrix and
$f^k=\langle \Psi_T|O^k (\Lambda - H)|\Psi_T \rangle$ is the known term; 
both $s^{l k}$ and $f^k$ are computed stochastically by a
Monte Carlo integration. 
These  linear equations (\ref{sys}) 
 are very similar to the ones introduced by Filippi and 
Fahy \cite{filippi} for the energy minimization of the Slater part.
In our formulation, there is no difficulty to optimize the Jastrow and the 
Slater part of the wavefunction at  the same time.

After the system (\ref{sys}) has been solved, we update the variational parameters
\begin{equation} \label{change}
\alpha_k=\alpha_k^{(0)} + \frac{\delta \alpha_k}{\delta \alpha_0}
\textrm{~~for~} k=1, \ldots, p
\end{equation}
and we obtain a new trial wavefunction $\Psi_T(\alpha)$. 
By repeating this iteration scheme several times, one approaches the 
convergence when $\frac{\delta\alpha_k}{\delta\alpha_0} \to 0$ for $k \ne 0$,
and in this limit  the SR conditions (\ref{srconditions}) implies 
the Euler equations of  the minimum energy. 
Obviously, the solution of the linear system (\ref{sys}) is affected 
by statistical errors, yielding statistical fluctuations of the 
final variational parameters $\alpha_k$ even when convergence has 
been reached, namely when  
the $\{ \alpha_k \}_{k=1,\ldots,p}$ fluctuate without drift around an average value.  
 We perform several iterations in that regime; in this way, the    
variational parameters can be determined more accurately 
by averaging them over all these iterations
and by evaluating also the corresponding  statistical error bars. 

It is worth noting  that the solution of the linear
system (\ref{sys}) depends on $\Lambda$ only through the $\delta \alpha_0$
variable. 
Therefore the constant $\Lambda$ indirectly controls the rate of change 
in the parameters at each step, i.e. the speed of the algorithm
 for convergence and the stability at equilibrium: a too small value will 
produce uncontrolled fluctuations for the variational parameters,
a too large one will allow convergence in an exceedingly large 
number of iterations. 
The choice of $\Lambda$ can be controlled by  evaluating the change 
of the wavefunction at each step as:
\begin{equation} 
\frac{ |\Psi^\prime_T-\Psi_T|^2 }{|\Psi_T|^2 } =\sum_{k,k^\prime>0} 
\delta \alpha_k ~ \delta\alpha_{k^\prime} ~ s^{kk^\prime}
\end{equation}

By keeping this value small enough during the optimization procedure, 
one can easily obtain a steady and stable convergence. 

Finally, we mention that 
the stochastic procedure is able in principle to perform
a global optimization, as discussed in Ref. \onlinecite{sorella} for the SR and in
Ref. \onlinecite{barba} for the Stochastic Gradient Approximation (SGA),
because the noise in the sampling can avoid the dynamics of the parameters to
get stuck into  local minima. 

\subsection{Variational and Diffusion Monte Carlo} 

We performed standard Variational (VMC) and Diffusion Monte Carlo (DMC), the latter within
the so called Fixed Node (FN) approximation,  which  allows to 
obtain the lowest energy state with the same nodes of a trial wavefunction. 
As trial state for FN, we have used  the  VMC wavefunction optimized with the
SR method described in the previous section.

\section{results}

\label{results}

We have carried out Quantum Monte Carlo calculations for atoms from $Li$ to
$P$, using the antisymmetrized geminal power times the Jastrow factor
($AGP+J$) to describe the atomic electronic structure. We
performed the optimization of both the geminal and the Jastrow part minimizing
the energy with the method described in Sec.\ref{method}. For all the atoms,
we considered  the minimal geminal expansion, 
corresponding to the $HF$ single determinant, and 
the simplest Jastrow factor with a single parameter (\ref{jastrows}), 
reported in Table \ref{slater}.
To improve the antisymmetric part, we increased  the number of
orbitals  in the geminal expansion, and for $Be$ and $Mg$ atoms 
we also systematically considered an improved Jastrow term, such as 
the three-body one described above (see Tables \ref{agpslater} and \ref{agpslater3}).
As one can notice from the Tables, our wavefunction parametrization is very
compact, even in the case of the mostly correlated states, 
since it contains always a relatively small number of parameters for each atom.

\begin{figure*}
%\centerline{
%\psfig{figure=vmcpercent.eps,width=12cm,angle=-90}}
\includegraphics[angle=-90,width=12cm]{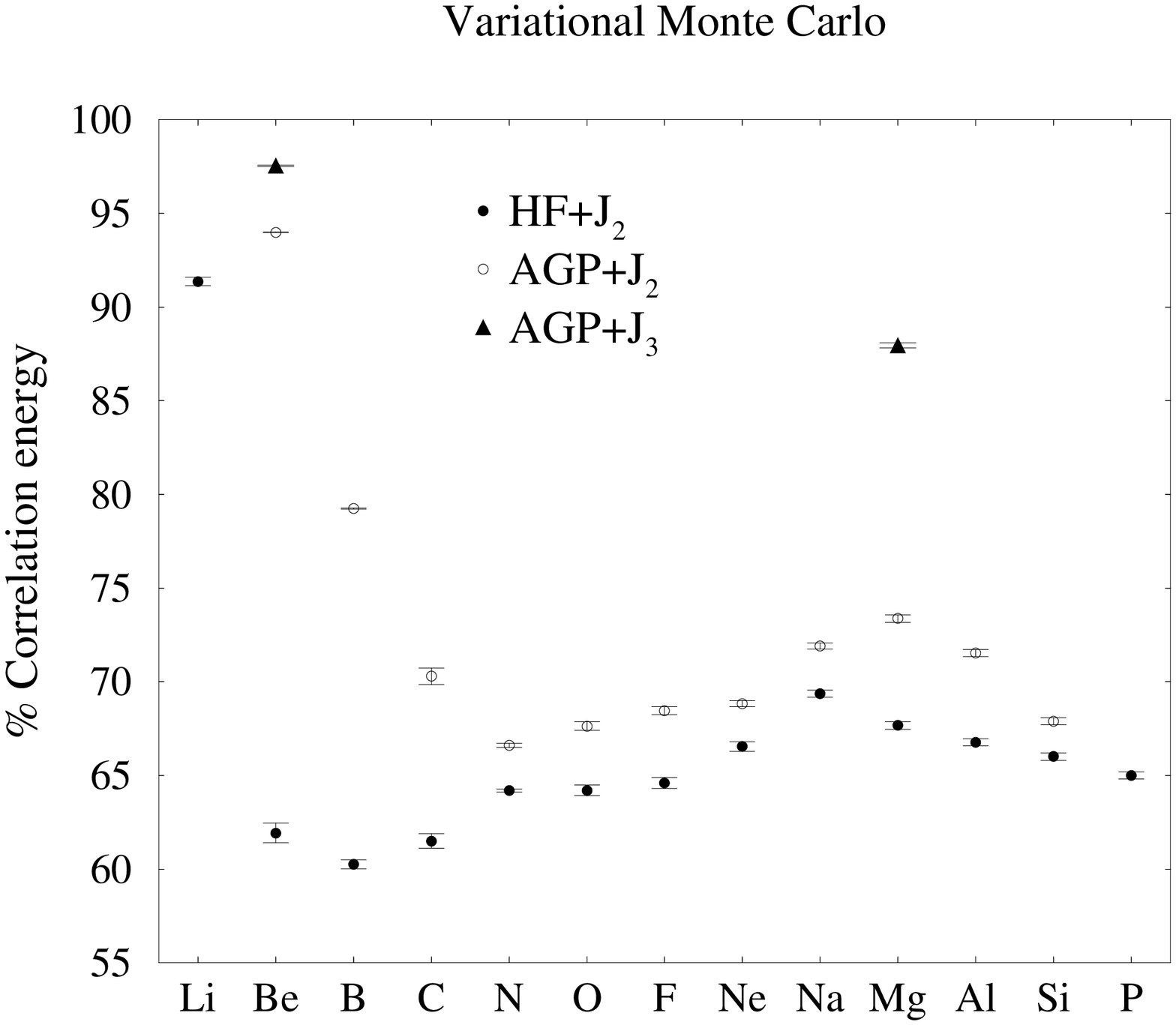}
\caption
{VMC correlation energies for $HF+J_2$ (minimal geminal expansion with a 
two-body Jastrow factor), $AGP+J_2$ (best geminal expansion)
and $AGP+J_3$ (best geminal with a three-body Jastrow factor)}
\label{vmc}
\end{figure*}

\begin{figure*}
%\centerline{
%\psfig{figure=fnpercent.eps,width=12cm,angle=-90}}
\includegraphics[angle=-90,width=12cm]{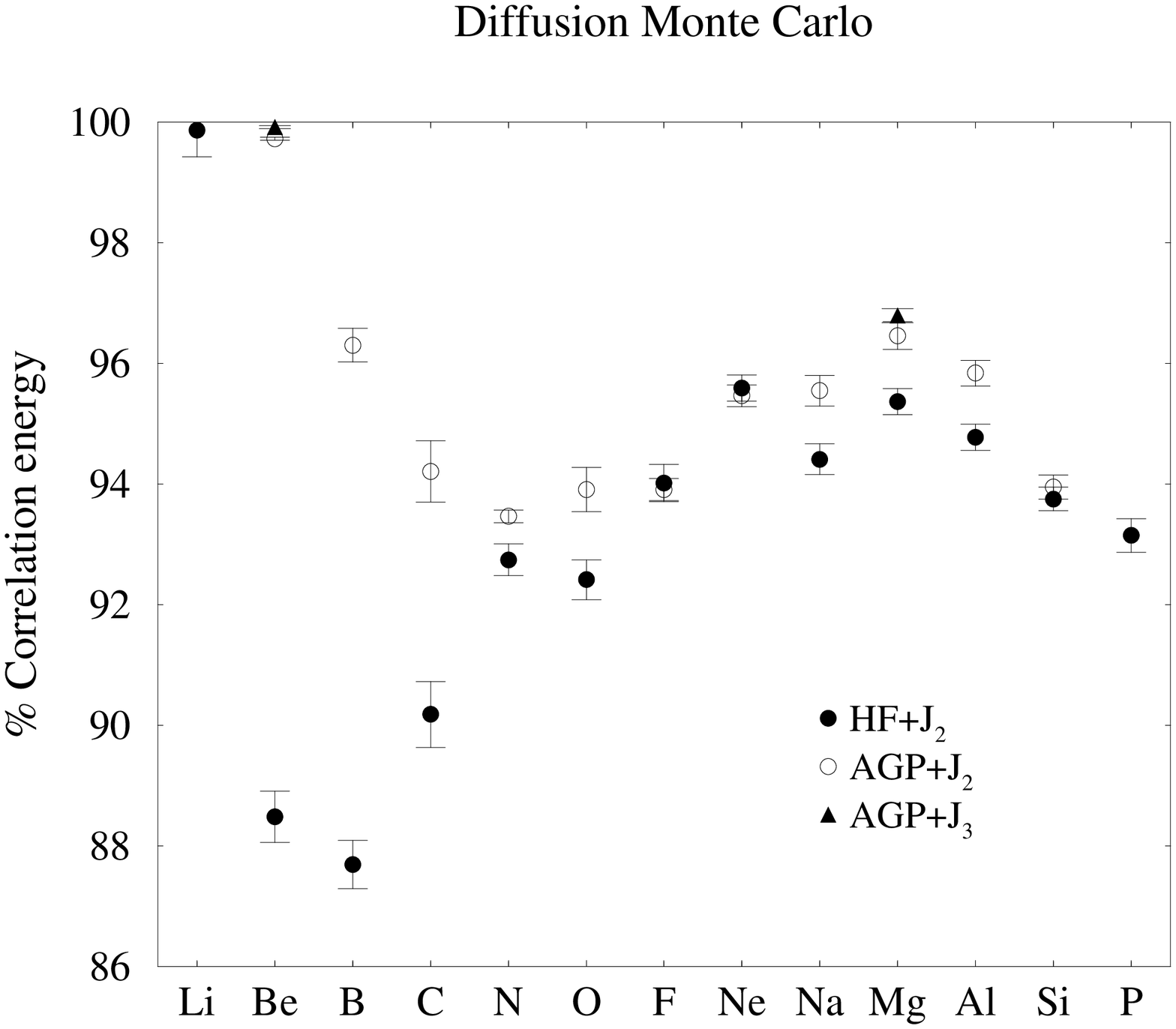}
\caption
{DMC correlation energies obtained by $HF+J_2$, $AGP+J_2$ and
$AGP+J_3$ wavefunctions} 
\label{dmc}
\end{figure*}

In order to judge the
outcome of our calculations, we computed the correlation energies and in
particular its fraction with respect to the exact ground state energy for the
nonrelativistic infinite nuclear mass Hamiltonian, estimated in Ref. \onlinecite{exact}.
The quality of the variational wavefunction can be seen by computing the
expectation value of the energy by means of the VMC calculations.
Furthermore, we carried out
DMC simulations within the FN approximation, which allows to optimize the amplitude
of the wavefunction inside each nodal volume, where its sign 
is given and fixed by the variational state. 
Therefore, the DMC energy depends on the quality of the nodal structure
of the variational wavefunction and the capability of improving the nodes
during the optimization is crucial to obtain  almost exact DMC energy
values. 
 To that purpose, it is very important to have a variational functional form
appropriate to reproduce the correct nodes. We show that the $AGP+J$
wavefunction satisfies well this requirement, yielding in all the atoms
studied here very good DMC results. 
The VMC and DMC energies are listed in Table \ref{energy};
in Figures \ref{vmc} and \ref{dmc} we plot the percentage of the correlation
energy recovered respectively by VMC and DMC calculations for different
atoms and wavefunctions. 

The VMC calculations with the minimal geminal expansion and the two body
Jastrow factor yield from $60 \%$ to $68 \%$ of the total
correlation energy, with the exception of the $Li$ atom, where $91.4 \%$ of the correlation energy
is obtained.  
Therefore, there is a sizable loss of accuracy in going from $Li$ to $Be$,
the worst case being the Boron atom. 
The corresponding DMC simulations get a large amount 
of the energy missing in the VMC calculations, recovering from $87.7\%$ to $99.9\%$ 
of the total correlation energy, but the dependence on the
atomic number shows the same behavior found in VMC: 
the worst results are obtained for $Be$, $B$ and $C$
atoms, due to the strong multiconfigurational nature of their ground states. As
well known, one can improve substantially the variational state of those atoms 
including not only the $1s^2 ~ 2s^2$ configuration but also the $2s ^2~ 2p^2$,
because of the near degeneracy of $2s$ and $2p$ orbitals. In this case the
$AGP+J$ ansatz is particularly efficient: by adding just one term in the geminal
expansion, we are able to remove this  loss of accuracy  in the correlation
energy both in the VMC and the DMC calculations.

In Table \ref{be}, we summarize some results obtained for $Be$ in previous
works and compare them with ours. $AGP$ calculations of atoms have been
performed only few times so far, the best one for $Be$ is reported in the last
row of the Table \ref{be}. Kurtz \textit{et al.} \cite{kurtz2}  were able to recover
$84 \%$ of correlation energy using a geminal expansion with a very large
basis; our variational $AGP+J$ wavefunction reaches $94 \%$ with a much
smaller basis ($1s$, $2s$ and $2p$ orbitals). 
By including a three-body Jastrow factor,
$97.5\%$ of the correlation energy is finally obtained, which is comparable 
with the best multiconfigurational wavefunctions previously studied \cite{huang}.
 
This outcome highlights the importance of
the Jastrow in reducing the geminal expansion and yielding a better
energy.
The nodal surface can be substantially improved 
with the present  approach,  
 because the pairing expansion contains implicitly not
only the four determinants $1s^2 2s^2$ and $1s^2 2p^2$, but also the other three
$2s^2 2p^2$ and six $2p^2 2p^2$, which can improve further the
wavefunction. Indeed, the geminal expansion reduces exactly to four determinants
in the limit $\lambda_{2s} \rightarrow 0$ and $\lambda_{2p} \rightarrow 0$
with constant ratio $\frac{\lambda_{2s}}{\lambda_{2p}}$. The fact that the
minimum energy is obtained for $\lambda_{2s} \ne 0$ and $\lambda_{2p} \ne 0$
(see Tables \ref{agpslater} and \ref{agpslater3}) clearly shows that the energy can be lowered by
considering the remaining configurations described above.
 Indeed  our  DMC energies are slightly {\em lower} than the 
ones by Huang {\it et al.}\cite{huang}, to our knowledge the 
best available ones  obtained with the four determinants: 
one $1s^2 2s^2$ and
three $1s^2 2p^2$. 
In order to determine accurate nodes for the corresponding DMC calculation
they used 
the two-body Jastrow factor similar to the one (\ref{jastrows}) we used 
or an highly involved three-body term much more complete than ours 
(for this reason our corresponding VMC energy is slightly worse in this case).     
We also verified, therefore, that a more accurate description of the Jastrow 
factor (which do not affect directly the nodes) is crucial to obtain better nodes, 
provided the variational parameters, belonging to both the Jastrow 
and the determinantal part, are optimized altogether.  
For instance, in
Ref. \onlinecite{sarsa} the authors optimized only the coefficients 
in front of the four determinants $1s 2s - 1s 2p$ 
and not the orbitals, obtaining for $Be$ energy not comparable with 
the best possible ones. 
The $AGP+J$ is simple enough to allow a feasible parametrization
of the variational state, by capturing the most important 
correlation.

We found that also $Mg$, $Al$ and $Si$, the second row atoms corresponding to
$Be$, $B$ and $C$ in the first row, have a 
quite strong multiconfigurational character, involving here
$3s$ and $3p$ orbitals. Analogously to the Beryllium case, for the $Mg$ we
have optimized both the two-body and the three-body Jastrow factor, together
with the $AGP$ wavefunction containing $3s$--$3p$ resonance. In this case,
although at the variational level the three-body wavefunction is much better
than the two-body one (see Fig. \ref{vmc}), 
that difference disappears almost completely in the DMC
results. This shows that the correction of the nodal surface allowed by the
more accurate three-body Jastrow does not seem to be crucial as in the $Be$
atom. On the other hand, the effect of the $AGP$ expansion is significant in
improving further the DMC calculation, which already yields good FN energies
even with a simple $HF+J$ trial wavefunction for atoms heavier than $C$ 
(percentage of  correlation energy  always greater than $92 \%$). By adding
the $3p$ contribution to the geminal we were able to recover $96.8 \%$ of the
correlation energy of $Mg$ (see Fig. \ref{dmc}). Also for $Al$ the
presence of the $3p$ orbital is significant in reducing the DMC energy, and
for $Si$ it seems important only in the VMC value.

Finally, by using the $AGP+J$ wavefunction, we optimized some atoms (from $N$ to $Na$) not
affected by nondynamic correlation; here, in order to obtain an improvement in
the VMC and in the DMC energies, we needed a bigger basis ($3s 2p 1d$) to be
used in the geminal expansion. 

\section{Conclusions and perspectives}

\label{conclusions}

In this work we have introduced a variational wavefunction 
which contains  the main ingredients of electron correlations:
the Jastrow factor, that allows to satisfy the electron--electron cusp 
condition, and the geminal expansion, that allows to consider  a 
  correlated multiconfiguration  
 state, with a numerically feasible scheme, namely 
by evaluating only a single but  appropriately defined determinant. 

 The application to atoms is particularly successful for low atomic number, 
where Hartree-Fock is particularly poor, due to the almost 
degenerate $2s-2p$ shells. 
 The case of Beryllium is an  interesting benchmark. 
Indeed,  by considering the change of the geminal part 
altogether with the Jastrow term, we  obtained   an excellent
representation of this correlated atom. Our results, presented in Table \ref{be}, 
are not only comparable but appear even better than the 
best multideterminantal schemes (using e.g. four Slater 
determinants), showing that 
it is possible to represent  non trivial correlated states 
by properly taking into account the 
interplay of the Jastrow term and the determinantal part of the wavefunction.
 Our variational energies for the other atoms (see Table \ref{energy}) can be
substantially lowered because we have considered, 
in this first application, a wavefunction with the two-body Jastrow factor.

As well known the variational energy of the 
Hartree-Fock wavefunction cannot be improved by extending
the variational  calculation  to a larger basis including 
all  particle-hole excitations applied to the Hartree-Fock state.
Analogously, the geminal wavefunction is not only stable with respect to these  
particle-hole configurations 
, but also to all possible states  
obtained by destroying a  singlet  pair on some orbital and 
creating another one  on another orbital.  
Though this wavefunction can take into account a big number of 
 configurations which may allow an  energy improvement, 
obviously it cannot include everything  within a single geminal,  
otherwise the wavefunction would be exact. 
Indeed there exist 
multiconfigurational states that are known  to be important for atoms 
like $C$ or those with larger $Z$ \cite{flad}, and that involve 
complicated multi--particle excitations to the Hartree-Fock state. These ones cannot 
be reduced to creation/destruction of singlet pairs and therefore cannot 
be handled with a single geminal function. 
However in our study we have found that the single geminal function 
with the proper Jastrow factor already provides satisfactory 
results for all atoms, yielding more than $93\%$ of the correlation energy 
in all cases studied by carrying out DMC simulations. 

The extension of this approach to molecules or more complex 
electronic systems is straightforward, and is indeed particularly promising. 
As pointed out in Ref. \onlinecite{rassolov}, the geminal wavefunction for a
diatomic system can correctly describe the interatomic Born--Oppenheimer  
potential from small to large 
distances, where, in this limit, 
 the wavefunction of isolated atoms can be smoothly recovered.
This important property cannot be 
satisfied  within the  Hartree-Fock theory,
even for the simplest $H_2$ molecule
(without contaminating the singlet ground state wavefunction).

 For an electronic system with  many atoms, the geminal expansion together 
with the Jastrow term is very similar to the so called 
Resonating Valence Bond (RVB) expansion, introduced by Pauling (see e.g. \cite{pauling})
long time ago, and revived recently 
by P.W. Anderson to consider the properties 
of strongly correlated electronic systems. The geminal part, when expanded 
in terms of Slater determinants, yields  a very  large and non trivial  
number of configurations, which increases exponentially with the number 
of atoms considered. The Jastrow factor in this case 
suppresses  the weight of those  configurations with two electrons 
close to the same atomic orbital, correctly describing the  
effect of  the  strong Coulomb repulsion.
We see therefore  the remarkable  advantage of this approach. Just for  
complex systems with many atoms an RVB wavefunction corresponding to an exponentially 
large number of configurations can be efficiently  used 
 for a more accurate description of electron correlation.  
Contrary to the conventional RVB expansion, it is appealing 
, not only  from the computational point of view, that these properties can 
be obtained by sampling a  single determinant wavefunction within  
  the Quantum Monte Carlo techniques.   

\begin{acknowledgments} 
 We gratefully acknowledge Saverio Moroni, for having provided us 
unpublished work on the Neon atom, and Claudia Filippi for useful 
comments on the preliminary version of the manuscript. We 
also thank G. Bachelet, F. Becca and C. Attaccalite for many useful 
discussions.  
This work was partially supported by MIUR--COFIN 2001.
\end{acknowledgments}

\appendix*

\section{Spin polarized geminal wavefunction}
\label{polarized}
In this appendix,  we consider the most general geminal wavefunction with 
definite spin $S=\frac{N_{\uparrow}-N_{\downarrow}}{2}$, where 
$N_\uparrow$ ($N_\downarrow $) is the number of spin--up (spin--down) electrons
and $N_{\uparrow} > N_{\downarrow}$ is assumed. 
To this purpose we introduce second quantized fermionic fields (see e.g. 
Fetter and Walecka \cite{fetter}) $\psi^{\dag} (\mathbf{r},\sigma) $ and $\psi
(\mathbf{r},\sigma)$, where $\mathbf{r}$ is the electron position and
$\sigma=\pm 1/2$ is its spin projection along the $z$--axis. These fields 
satisfy the canonical anticommutation rules:
\begin{equation} \label{ccr}
\left \{  \psi (\mathbf{r},\sigma) , \psi^{\dag} (\mathbf{r}^\prime,\sigma^\prime) 
\right\} = \delta_{\sigma \sigma^\prime} \delta (\mathbf{r} -\mathbf{r}^\prime).
\end{equation}

In these notations, the most general wavefunction with definite spin 
can be formally written in the following way:
\begin{equation} \label{pairing} 
|\Psi \rangle = P_N  \prod\limits_{ i =N_\downarrow+1}^{N_\uparrow} 
 \psi^{\dag}_{i,\uparrow}~ 
\exp ( \Phi^{\dag} )       |0 \rangle ,
\end{equation}
where $P_N$ is the projection on the given number of particles $N=
N_{\uparrow}+N_{\downarrow}$, 
$|0\rangle$ denotes the vacuum of electrons and  
 $\psi^{\dag}_{i,\uparrow}$  is the most 
generic (Bogoliubov) orbital with spin $S=1/2$: 
\begin{equation} \label{orbital} 
\psi^{\dag}_{i,\uparrow} = 
 \int \! d \mathbf{r} ~ \left (   \phi_i^{<}  (\mathbf{r}) \psi (\mathbf{r},\downarrow ) +
  \phi_i^{>}   (\mathbf{r}) \psi^{\dag }  (\mathbf{r},\uparrow ) \right ),
\end{equation}
 which is defined by the orbital functions $\phi_i^{>}$   for the creation 
of a particle with spin up and $ \phi_i^{<}$ for the 
annihilation of a particle with spin down. 
 For instance, a conventional Slater determinant of spin-up particles 
can be written as $\prod_{i} \psi^{\dag}_{i,\uparrow} |0 \rangle $, 
where $ \phi_i^{<}=0$. It is clear therefore  that this representation 
is more general and may provide a wavefunction $\Psi$ much more reach than 
the conventional Slater determinants.  

 Finally, the pairing creation operator $\Phi^{\dag}$ 
 is a singlet, namely $\exp ( \Phi^{\dag} ) |0 \rangle $ has spin zero, 
 and is defined by a generic symmetric 
 function $\Phi(\mathbf{r},\mathbf{r}^\prime)=\Phi(\mathbf{r}^\prime,\mathbf{r})$:
\begin{equation} \label{pair}
\Phi^{\dag} = \int d\mathbf{r} \int d\mathbf{r}^\prime  ~ \Phi(\mathbf{r}^{\prime},\mathbf{r}) \psi^{\dag} (\mathbf{r},\downarrow) 
\psi^{\dag} (\mathbf{r}^{\prime},\uparrow) .
\end{equation}

Our purpose is to show here that the value of the wavefunction 
$\Psi$ can be simply  computed, 
similarly to a conventional Slater determinant,  on 
each configuration
$x=\{\mathbf{r}_{1,\uparrow},\ldots,\mathbf{r}_{N_\downarrow,\downarrow}\}$,
where $\mathbf{r}_{i,\uparrow}$ are  the positions  of spin--up particles and
$\mathbf{r}_{i,\downarrow}$ are the  spin--down ones.
These configurations can be generally written as:
\begin{equation} \label{conf}
 \langle x | = 
\langle 0 | \prod\limits_{i=1}^{N_\uparrow}  \psi (\mathbf{r}_i,\uparrow)  
  \prod\limits_{j=1}^{N_\downarrow}  \psi (\mathbf{r}_j,\downarrow).
\end{equation}
 Indeed the value $F$ of the wavefunction on $\langle x |$ is:
\begin{widetext}
\begin{equation} \label{simpledet}
 F  =  \langle x | \Psi \rangle  
    =  \left \langle  0  \left |  \prod_i \psi(\mathbf{r}_i,\uparrow) 
\prod_j \psi(\mathbf{r}_j,\downarrow)  \prod\limits_{k=N_\downarrow+1}^{N_\uparrow} 
\psi^{\dag}_{k,\uparrow} ~ \exp ( \Phi^{\dag} ) \right | 0 \right \rangle .
\end{equation}
\end{widetext}
Now we insert the identity $\exp (-\Phi^\dag) \exp (\Phi^\dag)$ between each
fermionic field in the above equation (\ref{simpledet}):
\begin{eqnarray}
F & = & \left \langle 0 \left | \exp (\Phi^\dag) \exp (-\Phi^\dag) \psi
(\mathbf{r}_1,\uparrow) \exp (\Phi^\dag) \cdots \right . \right . \nonumber \\
& & \left . \left . \cdots \exp (-\Phi^\dag) \psi
(\mathbf{r}_{N_\downarrow},\downarrow)  \exp (\Phi^\dag) \cdots
\right
. \right .\nonumber \\
& & \left . \left . \cdots \exp(-\Phi^\dag) \psi^\dag_{N_\uparrow,\uparrow} \exp ( \Phi^{\dag}
) \right | 0 \right \rangle .  \label{inter}
\end{eqnarray}
Exploiting the relation valid for generic operators $A$, $B$ and $C$:
\begin{equation}
\label{comm}
\exp (-A) ~B~ \exp (A) = B - \left [ A,B \right ] + \frac{1}{2} \left [ A,
\left [ A, B\right ] \right ] + \ldots 
\end{equation}
one is able to evaluate the following terms:
\begin{eqnarray}
\exp (- \Phi^{\dag} ) \psi (\mathbf{r}_i, \uparrow) \exp ( \Phi^\dag ) & = & 
\nonumber \\
\psi (\mathbf{r}_i, \uparrow) - \int \! d \mathbf{r} ~ \Phi( \mathbf{r}_{i,\uparrow},
\mathbf{r}) \psi^\dag (\mathbf{r}, \downarrow) & & \nonumber \\
\exp (- \Phi^{\dag} ) \psi (\mathbf{r}_i, \downarrow) \exp ( \Phi^\dag ) & = &
\nonumber \\
\psi (\mathbf{r}_i, \downarrow) + \int \! d \mathbf{r} ~ \Phi( \mathbf{r},
\mathbf{r}_{i,\downarrow}) \psi^\dag (\mathbf{r}, \uparrow) & & \nonumber \\
\exp (- \Phi^{\dag} ) \psi^\dag_{i,\uparrow} \exp ( \Phi^\dag ) & = &
\nonumber \\
 \psi^\dag_{i,\uparrow} + \int \! d \mathbf{r} \int \! d \mathbf{r}^\prime
~\Phi(\mathbf{r}, \mathbf{r}^\prime) \phi_i^{<} (\mathbf{r}^\prime)
\psi^\dag(\mathbf{r}, \uparrow)
\end{eqnarray}
In order to derive the above relations, notice that all the terms in the RHS
of Eq. \ref{comm} are always zero beyond the first two.
After substituting  the expressions in Eq. \ref{inter} and by using 
$\langle 0 | \exp ( \Phi^{\dag} )  = \langle 0 | $, 
$\psi (\mathbf{r},\sigma ) |0\rangle =0$ and $\langle 0 |\psi^{\dag}  (\mathbf{r},\sigma) =0 $, 
one can iteratively apply the canonical commutation rules (\ref{ccr}) 
and a simplified result for $F$ is obtained:
\begin{equation}
F= \left \langle 0 \left | \prod\limits_{i=1}^{N_\uparrow} \psi(\mathbf{r}_i,\uparrow) 
\prod\limits_{i=1}^{N_\uparrow} \tilde \psi^{\dag}_{i,\uparrow} \right |0 \right
\rangle 
\end{equation}
where  $\tilde \psi^\dag_{i,\uparrow}$ is the creator of an orbital function
of the type (\ref{orbital}), with transformed orbitals:
\begin{subequations}
\begin{eqnarray} 
\phi_i (\mathbf{r}) &=& \Phi(\mathbf{r},\mathbf{r}_{i,\downarrow}) 
\label{pippo1}  \\
& & {\rm for ~ }i=1,\cdots , N_\downarrow \nonumber \\
\bar  \phi_i (\mathbf{r}) &=&  \phi_i^{>}(\mathbf{r}) + \int \!
d\mathbf{r}^\prime ~ \Phi(\mathbf{r},\mathbf{r}^\prime) \phi_i^{<}
 (\mathbf{r}^\prime) 
\label{pippo2} \\
& & {\rm for~ } i=N_\downarrow+1, \cdots,  N_\uparrow  \nonumber 
\end{eqnarray} 
\end{subequations}
 Then the final value of $F$ can be simply computed by a {\em single} 
determinant, as it represents just the value of a  $N_\uparrow  \times
N_\uparrow$ 
Slater determinant with 
orbitals given in (\ref{pippo1}) and (\ref{pippo2}) on the spin-up configurations, yielding
the final expression (\ref{finaldet}) reported in the text.

\begin{table*}[!hbp]
{ \small
\caption{\label{energy}Total energies in variational ($E_{VMC}$) and 
diffusion ($E_{DMC}$) Monte Carlo calculations; 
the percentages of correlation energy recovered in
VMC ($E^{VMC}_c(\%)$) and DMC ($E^{DMC}_c(\%)$) have been evaluated using
the exact ($E_0$) and Hartree--Fock ($E_{HF}$) energies from
Ref.\onlinecite{exact}. $M$ is the number of terms in the pairing expansion.
The energies are in \emph{Hartree}.} 
\begin{ruledtabular}
\begin{tabular}{l c d d d d d d }
& $M$ & \makebox[0pt][c]{$E_0$} & \makebox[0pt][c]{$E_{HF}$} &
\makebox[0pt][c]{$E_{VMC}$} & \makebox[0pt][c]{$E^{VMC}_c(\%)$} 
& \makebox[0pt][c]{$E_{DMC}$} & \makebox[0pt][c]{$E^{DMC}_c(\%)$} \\
\hline
Li & 1 & -7.47806 & -7.432727 & -7.47415(10)    & 91.4(2) & -7.4780(2)   & 99.9(4) \\
Be & 2 & -14.66736 & -14.573023 & -14.63145(5)   & 61.9(5) & -14.6565(4) & 88.5(4) \\
   & 5 &  &  & -14.661695(10) & 93.995(11) &  
-14.66711(3) &  99.73(3) \\           
   & 5\footnotemark[1] &  &  & -14.66504(4) & 97.54(5) & 
-14.66728(2) &  99.92(2) \\
B  & 2 & -24.65391 & -24.529061 & -24.6042(3)    & 60.3(2) & -24.63855(5)  & 87.7(4) \\
   & 5 & & & -24.62801(4) & 79.25(4) & 
-24.6493(3)   & 96.3(3) \\ 
C  & 2 & -37.8450  & -37.688619 & -37.7848(6)   & 61.5(4) & -37.8296(8)  & 90.2(5) \\
   & 5 & & & -37.7985(7)  & 70.3(4) &
 -37.8359(8) &  94.2(5)\\
N  & 2 & -54.5892 & -54.400934 & -54.52180(15)  & 64.20(8) & -54.57555(5) & 92.7(3) \\
   & 2\footnotemark[2]  &   &   & -54.52565(15)  & 66.20(8) & -54.5753(4)  & 92.6(2) \\
   & 14 &          &            &  -54.5263(2)  & 66.62(11) & -54.5769(2) & 93.47(10) \\
O  & 3 & -75.0673 & -74.809398 & -74.9750(7) & 64.2(3) & -75.0477(8) & 92.4(3) \\
   & 14 &          &             & -74.9838(6) & 67.6(2) & -75.0516(9) & 93.9(3) \\
F  & 4 & -99.7339 & -99.409349 & -99.6190(8) & 64.6(3) & -99.7145(15) & 94.0(5) \\
   & 14 &          &           & -99.6315(7) & 68.4(2) & -99.7141(6) & 93.91(18) \\
Ne & 5 & -128.9376 & -128.547098 & -128.8070(10)   & 66.6(3) & -128.9204(8) & 95.6(2) \\
   & 14 &          &             &  -128.8159(6) & 68.83(17) & -128.9199(7) & 95.47(18) \\
Na & 5 & -162.2546 & -161.858912 & -162.1334(7) & 69.37(19) & -162.2325(10) & 94.4(2) \\
   & 9 &           &             & -162.1434(7) & 71.91(16) & -162.2370(10) & 95.5(2) \\
Mg & 6 & -200.053  & -199.614636 & -199.9113(8) & 67.67(19)  & -200.0327(9) & 95.4(2) \\
   & 9 & & & -199.9363(8) & 73.38(19)     &  
-200.0375(10) &  96.5(2) \\
   & 9\footnotemark[1] & & & -200.0002(5) & 87.95(12) &
-200.0389(5) & 96.79(11) \\  
Al & 6 & -242.346  & -241.876707 & -242.1900(9) & 66.77(19)  & -242.3215(10) & 94.8(2) \\
   & 9 & & & -242.2124(9) & 71.53(19)  & 
-242.3265(10) &  95.8(2) \\
Si & 6 & -289.359  & -288.854363 & -283.1875(10)   & 66.0(2)  &  -289.3275(10) & 93.8(2) \\
   & 9 & &  & -289.1970(10) & 67.9(2) &   
 -289.3285(10) & 94.0(2) \\
P  & 6 & -341.259  & -340.718781 & -341.0700(10)   & 65.0(2)  &  -341.2220(15)  & 93.2(3) \\
\end{tabular}
\footnotetext[1]{Wavefunction with a three body Jastrow factor.}
\footnotetext[2]{Wavefunction with a two body Jastrow factor spin contaminated.}
\end{ruledtabular}
}
\end{table*}

\begin{table*}[!hbp]
{ \small
\caption{\label{be} Comparison of the energies obtained by various authors for $Be$.}
\begin{ruledtabular}
\begin{tabular}{l d d  d d }
& \multicolumn{1}{c}{basis} & \multicolumn{1}{c}{Jastrow} 
& \multicolumn{1}{c}{VMC} & \multicolumn{1}{c}{DMC} \\
\hline
\multicolumn{1}{c}{present work} 
& \multicolumn{1}{c}{2s1p} & \multicolumn{1}{c}{two body} & -14.661695(10) & -14.66711(3) \\
\multicolumn{1}{c}{Huang \textit{et al.}\cite{huang}} 
& \multicolumn{1}{c}{2s1p} &  \multicolumn{1}{c}{two body} & -14.66088(5) & -14.66689(4) \\
\multicolumn{1}{c}{present work} 
& \multicolumn{1}{c}{2s1p} & \multicolumn{1}{c}{three body} & -14.66504(4) &  -14.66728(2) \\
\multicolumn{1}{c}{Huang \textit{ETA.}\cite{huang}} 
& \multicolumn{1}{c}{2s1p} & \multicolumn{1}{c}{three body} &  -14.66662(1) &  -14.66723(1) \\
\multicolumn{1}{c}{Sarsa \textit{et al.}\cite{sarsa}} 
& \multicolumn{1}{c}{2s1p} & \multicolumn{1}{c}{three body} & -14.6523(1) & \\
\multicolumn{1}{c}{Kurtz \textit{et al.}\cite{kurtz2}} 
& \multicolumn{1}{c}{6s3p2d} & \multicolumn{1}{c}{none} & -14.6171 &  \\
\end{tabular}
\end{ruledtabular}
}
\end{table*}

\begin{table*}[!hbp]
{ \small
\caption{\label{slater} \textbf{HF+J (two body) wavefunctions} \newline
Parameters of the Jastrow and the pairing function with the notation described
in the text. ``\#'' means that the corresponding parameter has to be evaluated 
from the cusp condition in Eq. \ref{cusp2}. 
The line over the orbitals label refers to the unpaired ones. 
The values are given with the statistic error due to the stochastic approach in
the minimization.}
\begin{ruledtabular}
\begin{tabular}{l | d | r d d d d }
& \multicolumn{1}{c}{$b$} & orbital &
\makebox[0pt][c]{$Z_1$} & \makebox[0pt][c]{$Z_2$}
& \makebox[0pt][c]{$p$}   & \multicolumn{1}{c}{$\lambda$} \\ 
\hline
Li & 0.731(3) & $1s$ & 3.556(2)    & 2.3741(15)   &  \mbox{\#} & ~1.0  \\
   & & $\overline{2s}$ & 1.4289(12)   & 0.5380(2)  & \mbox{\#}      &     \\
\hline
Be & 0.773(2) & $1s$ & 4.569(6)    & 3.289(5)   &   \mbox{\#}  & ~1.0  \\
   & & $2s$ & 2.602(3)   & 0.7850(8)   &   \mbox{\#}  & ~1.0 \\
\hline
B & 0.877(2) & $1s$ & 5.569(5)    & 4.195(4)   &  \mbox{\#}  & ~1.0  \\
  & & $2s$ & 3.527(2)   & 1.0788(2)  & \mbox{\#}  & ~1.0 \\
  & & $\overline{2p}$ & 2.437(4)   & 1.1001(5)  &   ~0.2664(3) &   \\  
\hline
C & 0.990(2) & $1s$ & 6.533(6)    & 5.075(8)   &  \mbox{\#}  & ~1.0 \\
  & & $2s$ & 4.475(3)   & 1.3552(3)  &   \mbox{\#}  & ~1.0 \\
  & & $\overline{2p}$ & 2.9835(12)   & 1.3886(5)  &   ~0.2374(4) &      \\  
\hline
N & 1.110(3) & $1s$ & 7.461(8)    & 5.901(15)   & \mbox{\#}  & ~1.0 \\
  & & $2s$ & 5.387(3)   & 1.6323(1)  &  \mbox{\#}  & ~1.0 \\
  & & $\overline{2p}$   & 3.4908(9)   & 1.6373(4)  &  ~0.2036(3) &\\
\hline
N & \makebox[0pt][c]{contaminated} 
& $1s$ & 7.2738(17) & 5.522(7) & \mbox{\#}  & ~1.0 \\
  & \makebox[0pt][c]{$b_{\uparrow \downarrow}$=0.940(3)}   
  & $2s$ & 5.254(2)   & 1.6692(15)  &  \mbox{\#}  & ~1.0 \\
  & \makebox[0pt][c]{$b_{\uparrow \uparrow}$=0.624(4)}
  & $\overline{2p}$   & 3.4544(4)   & 1.66559(13)  &  ~0.21593(9) &\\
\hline
O & 1.152(2) & $1s$ & 8.303(5)    & 6.53(3)   & \mbox{\#}  & ~1.0 \\
  & & $2s$ & 6.428(2)   & 1.9362(5)  &  \mbox{\#}  & ~1.0 \\
  & & $2p$ & 3.9921(7)   & 1.84932(15)  &  ~0.1648(2)  & ~1.0 \\
  & & $\overline{2p}$   & 3.9921(7)   & 1.84932(15)  &  ~0.1648(2)  & \\
\hline
F & 1.226(9) & $1s$ & 9.171(3)    & 6.82(2)   & \mbox{\#}  & ~1.0\\
  & & $2s$ & 7.380(2)   & 2.2402(4)  &  \mbox{\#}  & ~1.0 \\
  & & $2p$ & 4.5025(2)   & 2.0758(2)  &  ~0.1451(10)  & ~1.0 \\
  & & $\overline{2p}$  & 4.5025(2)   & 2.0758(2)  &  ~0.1451(10)  &\\
\hline
Ne & 1.321(3) & $1s$ & 10.103(4)    & 7.01(4)   & \mbox{\#}  & ~1.0  \\
   & & $2s$ & 8.341(10)   & 2.5319(5)  &   \mbox{\#}  & ~1.0 \\
   & & $2p$ & 5.0273(10)   & 2.3155(4)  &  ~0.1358(2)  & ~1.0\\
\hline
Na & 1.514(7) & $1s$ & 11.102(2)    & 7.58(5)   & \mbox{\#}  & ~1.0  \\
   & & $2s$ & 8.823(2)   & 2.8795(4)  &  \mbox{\#}  & ~1.0 \\
   & & $2p$ & 5.7178(9)   & 2.8092(8)  &  ~0.18027(18) & ~1.0 \\
   & & $\overline{3s}$ & 1.540(3)  &  0.734(2)   & ~0.1265(2)  &\\
\hline
Mg & 1.654(6) & $1s$ & 12.0855(15)    & 7.96(5)   & \mbox{\#}  & ~1.0 \\
   & & $2s$ & 9.349(4)   & 3.2611(16)  &  \mbox{\#}  & ~1.0 \\
   & & $2p$ & 6.3795(10)   & 3.2905(5)  &  ~0.21569(17)  & ~1.0 \\
   & & $3s$ & 1.9288(10)   & 1.02214(13)  &  ~0.1449(2)  & ~1.0 \\
\hline
Al & 1.812(8) & $1s$ & 13.0791(17)    & 8.43(7)   & \mbox{\#} & ~1.0  \\
   & & $2s$ & 9.846(8)   & 3.646(2)  &  \mbox{\#}  & ~1.0 \\
   & & $2p$ & 7.0560(15)   & 3.7873(7)  &  ~0.2597(3)  & ~1.0 \\
   & & $3s$ & 2.24(8)  & 1.2627(2) &  ~0.1445(2)  & ~1.0 \\
   & & $\overline{3p}$ & 1.83(2)   & 0.8866(2)  &  ~0.1012(7)  &\\
\hline
Si & 1.961(7) & $1s$ & 14.072(2)    & 8.77(8)   & \mbox{\#}  & ~1.0  \\
   & & $2s$ & 10.40(5)   & 4.0275(8)  &  \mbox{\#}  & ~1.0\\
   & & $2p$ & 7.703(10)   & 4.261(5)  &  ~0.2944(12)  & ~1.0 \\
   & & $3s$ & 2.468(5) & 1.46(2)   &  ~0.1241(2)  & ~1.0 \\
   & & $\overline{3p}$ & 2.274(2) & 1.16(2)  &  ~0.1227(3)  & \\
\hline
P & 2.074(9) & $1s$ & 15.053(4)    & 9.10(15)   & \mbox{\#}  & ~1.0 \\
  & & $2s$ & 10.997(5)   & 4.4269(5)  &  \mbox{\#}  & ~1.0 \\
  & & $2p$ & 8.346(2)   & 4.7519(12)  &  ~0.3273(17)  & ~1.0 \\
  & & $3s$ & 2.7386(5) & 1.62(2)   &  ~0.1129(3) & ~1.0 \\
  & & $\overline{3p}$ & 2.5918(18) & 1.3479(7)  &  ~0.1125(2)  &  \\
\end{tabular}
\end{ruledtabular}
}
\end{table*}

\begin{table*}[!hbp]
{ \small
\caption{\label{agpslater}\textbf{AGP+J (two body) wavefunctions} \newline
The notations are the same as in Table \ref{slater}.}
\begin{ruledtabular}
\begin{tabular}{l | d | r d d d d }
& \multicolumn{1}{c}{$b$} & orbital &
\makebox[0pt][c]{$Z_1$} & \makebox[0pt][c]{$Z_2$}
& \makebox[0pt][c]{$p$}   & \multicolumn{1}{c}{$\lambda$}\\ 
\hline
Be & 1.009(5)  & $1s$ & 4.7139(6)    & 3.3118(3)   &  \mbox{\#} & ~1.0  \\
   & & $2s$ & 2.27561(16)   & 0.78955(2)  &  \mbox{\#} & -1.95(8)~\mbox{$10^{-3}$} \\
   & & $2p$ & 3.389(2)    & 1.06495(4)  &  0.8878(18) & ~3.49(15)~\mbox{$10^{-4}$} \\
\hline
B & 1.005(7) & $1s$ & 5.6509(8)    & 4.2041(7)   &  \mbox{\#} & ~1.0  \\
  & & $2s$ & 3.35545(19)   & 1.07294(2)  &  \mbox{\#} & -4.41(7)~\mbox{$10^{-3}$} \\
  & & $2p$ & 3.701(7)   & 1.46298(13)  &   ~0.827(2) & 6.92(10)~\mbox{$10^{-4}$}  \\  
  & & $\overline{2p}$ & 2.4093(3)   & 1.07197(4)  &   ~0.28102(8)   & \\  
\hline
C & 1.036(5) & $1s$ & 6.5420(6)    & 5.0494(8)   &   \mbox{\#} & ~1.0   \\
  & & $2s$ & 4.4094(3)   & 1.353943(18)  &  \mbox{\#} & -5.40(10)~\mbox{$10^{-3}$} \\
  & & $2p$ & 4.289(3)   & 1.8817(2)  &   ~0.7769(8) & 7.53(14)~\mbox{$10^{-4}$}  \\  
  & & $\overline{2p}$ & 2.95758(8)   & 1.36288(2)  &   ~0.23276(2)   & \\  
\hline
N & 1.124(5) & $1s$ & 7.4553(11)    & 5.866(2)   &   \mbox{\#} & ~1.0   \\
  & & $2s$ & 5.5031(12)   & 1.61305(8)  &  \mbox{\#} & -2.345(9)~\mbox{$10^{-2}$} \\
  & & $2p$ & 5.708(9)   & 2.5488(8)  &   ~0.9834(14) & 2.538(9)~\mbox{$10^{-3}$}  \\ 
  & & $3s$ & 3.415(2)   &    &    0.0 \footnotemark[1] & -7.521(4)~\mbox{$10^{-4}$} \\ 
  & & $3p$ & 2.1914(6)  &    &    0.0 \footnotemark[1] & 3.454(13)~\mbox{$10^{-4}$} \\
  & & $3d$ & 2.6498(6)  &    &    0.0  \footnotemark[1] & 1.881(4)~\mbox{$10^{-4}$} \\
  & & $\overline{2p}$ & 3.48977(11)   & 1.63502(2)  &   ~0.205373(12)   & \\  
\hline
O & 1.2073(10) & $1s$ & 8.3359(4)    & 6.5514(15)   &   \mbox{\#} & ~1.0   \\
  & & $2s$ & 6.4248(3)   & 1.94127(5)  &  \mbox{\#} & -9.54(6)~\mbox{$10^{-3}$} \\
  & & $2p$ & 3.97418(5)  & 1.77990(4) &  ~0.168724(8) & -5.73(4)~\mbox{$10^{-3}$}  \\ 
  & & $3s$ & 5.3337(16)   &    &    0.0 \footnotemark[1] & 3.09(2)~\mbox{$10^{-4}$} \\ 
  & & $3p$ & 2.00849(8)  &    &    0.0 \footnotemark[1] & 2.156(13)~\mbox{$10^{-4}$} \\
  & & $3d$ & 2.8598(5)  &    &    0.0  \footnotemark[1] & 5.02(3)~\mbox{$10^{-5}$} \\
  & & $\overline{2p}$ & 3.98928(8)   & 1.85965(3)  &   ~0.172589(8)   & \\  
\hline
F & 1.310(3) & $1s$ & 9.232(2)    & 7.037(14)   &   \mbox{\#} & ~1.0   \\
  & & $2s$ & 7.0901(19)   & 2.1520(2)  &  \mbox{\#} & -1.30(6)~\mbox{$10^{-2}$} \\
  & & $2p$ & 4.4857(4)  & 2.00946(14) &  ~0.15308(6) & -9.15(19)~\mbox{$10^{-3}$}  \\ 
  & & $3s$ & 3.0007(7)   &    &    0.0 \footnotemark[1] & 7.92(15)~\mbox{$10^{-4}$} \\ 
  & & $3p$ & 2.2966(5)  &    &    0.0 \footnotemark[1] & 3.25(13)~\mbox{$10^{-4}$} \\
  & & $3d$ & 3.2500(10)  &    &    0.0  \footnotemark[1] & 6.51(18)~\mbox{$10^{-5}$} \\
  & & $\overline{2p}$ & 4.5034(5)   & 2.09241(15)  &   ~0.15589(4)   & \\  
\hline
Ne & 1.3500(10) & $1s$ & 10.1233(3)    & 7.319(4)   &   \mbox{\#} & ~1.0   \\
  & & $2s$ & 8.0868(2)   & 2.47046(2)  &  \mbox{\#} & 4.650(18)~\mbox{$10^{-2}$} \\
  & & $2p$ & 5.00752(4)   & 2.28873(2)  &  ~0.133639(5) & 3.028(13)~\mbox{$10^{-2}$}  \\ 
  & & $3s$ & 3.65736(11)   &     &   0.0  \footnotemark[1] & -2.668(11)~\mbox{$10^{-3}$} \\ 
  & & $3p$ & 1.74505(5)  &  &    0.0   \footnotemark[1] & -1.379(7)~\mbox{$10^{-4}$} \\
  & & $3d$ & 2.74243(12)  & &    0.0  \footnotemark[1]  & -1.102(7)~\mbox{$10^{-4}$} \\
\hline
Na & 1.5469(11) & $1s$ & 11.1214(11)    & 7.807(19)   &   \mbox{\#} & ~1.0   \\
  & & $2s$ & 8.6638(18)   & 2.82430(13)  &  \mbox{\#} & -8.43(19)~\mbox{$10^{-3}$} \\
  & & $2p$ & 5.7308(3)   & 2.78883(12)  &  ~0.19053(5) & -5.69(17)~\mbox{$10^{-3}$}  \\ 
  & & $3s$ & 3.8585(14)   &     &   0.0  \footnotemark[1] & 2.86(12)~\mbox{$10^{-4}$} \\ 
  & & $3p$ & 3.1693(8)  &  &    0.0   \footnotemark[1] & 1.26(8)~\mbox{$10^{-4}$} \\
  & & $\overline{3s}$ & 1.53832(2)  & 0.7536(16) & 0.135076(3)  &  \\
\hline
Mg & 1.692(6) & $1s$ & 12.0983(14)    & 8.12(4)   &   \mbox{\#} & ~1.0 \\
   & & $2s$ & 9.3103(19)   & 3.2525(3)  &   \mbox{\#}  & ~1.0 \footnotemark[2]\\
   & & $2p$ & 6.3862(5)   & 3.28864(16)  &   ~0.22040(7) & ~1.0 \footnotemark[2] \\
   & & $3s$  &  1.7571(2)  & 0.96373(13) & 0.11295(7) & 1.986(10)~\mbox{$10^{-4}$}  \\
   & & $3p$  & 3.417(15)  & 1.2284(2) & 0.439(11) &  -2.825(15)~\mbox{$10^{-5}$}  \\ 
\hline
Al & 1.840(8) & $1s$ & 13.084(2)    & 8.47(9)   &   \mbox{\#}  & ~1.0  \\
   & & $2s$ & 9.826(8)   & 3.6344(12)  &  \mbox{\#}  & ~1.0 \footnotemark[2]\\
   & & $2p$ & 7.0532(16)   & 3.7780(8)  &   ~0.2597(3) & ~1.0  \footnotemark[2]  \\
  & & $3s$   & 2.190(3) & 1.2568(11)  &  0.1487(6) & 7.781(13)~\mbox{$10^{-4}$}  \\
  & & $3p$   & 5.58(2)  & 1.5531(7)   & ~0.1416(7) & -1.004(10)~\mbox{$10^{-4}$}  \\ 
  & & $\overline{3p}$ & 1.916(3)   & 0.9374(6)  &   ~0.1282(5)  & \\  
\hline
Si & 1.968(3) & $1s$ & 14.0697(7)    & 8.68(4)   &   \mbox{\#} & ~1.0  \\
   & & $2s$ & 10.4077(18)   & 4.0249(4)  &   \mbox{\#}  & ~1.0   \footnotemark[2] \\
   & & $2p$ & 7.7030(7)   & 4.2614(3)  &   ~0.29444(12) & ~1.0  \footnotemark[2]   \\
   & & $3s$  & 2.4437(3)  & 1.42785(13) &  0.12038(3) & 7.327(16)~\mbox{$10^{-4}$}  \\
   & & $3p$   & 6.110(12)  & 1.8471(3)   & ~0.19670(14) & -8.961(18)~\mbox{$10^{-5}$}  \\ 
   & & $\overline{3p}$ & 2.2614(8) & 1.16(1)    &   ~0.1237(2)  & \\  
\end{tabular}
%\footnotetext[1]{With the three body jastrow factor}
\footnotetext[1]{In the case of dynamic correlation, we used single zeta
orbitals beyond the $HF$ ones.}
\footnotetext[2]{For $Mg$, $Al$ and $Si$, we optimized only the $\lambda$'s
related to the $3s -3p$ shells, whose contribution is 
the most important in describing the resonance in that multiconfigurational ground states.}
\end{ruledtabular}
}
\end{table*}

\begin{table*}[!hbp]
{ \small
\caption{\label{agpslater3}\textbf{AGP+J (three body) wavefunctions} \newline
The notations are the same as in Table \ref{slater}. The parametrization for
the three-body Jastrow factor is also reported. }
%\begin{ruledtabular}
\begin{tabular}{l | d d d d d }
\multicolumn{1}{l}{$Be$} & & & & & \\
\hline
\hline
geminal &
\makebox[0pt][c]{$Z_1$} & \makebox[0pt][c]{$Z_2$}
& \makebox[0pt][c]{$p$}   & \multicolumn{1}{c}{$\lambda$} & \\ 
\hline
\multicolumn{1}{r|}{$1s$} & 7.48(8)  & 3.545(2)  & \mbox{\#} & ~1.0 &  \\
\multicolumn{1}{r|}{$2s$} & 2.403(3) & 0.8181(3) & \mbox{\#} &
3.95(9)~\mbox{$10^{-3}$} & \\
\multicolumn{1}{r|}{$2p$} & 2.86(5)  & 1.1064(9) & ~1.245(9)
&-6.27(9)~\mbox{$10^{-4}$} & \\  
\hline
\hline
Jastrow 2 body & \multicolumn{1}{c}{$b$} 
& \multicolumn{4}{c}{} \\
\hline
  &  0.7935(19) & & & & \\
\hline
\hline
Jastrow 3 body &  
 \makebox[0pt][c]{$Z_1$} & \makebox[0pt][c]{$Z_2$}
& \makebox[0pt][c]{$a_0$}   & \multicolumn{1}{c}{$a_1$} 
& \multicolumn{1}{c}{$a_2$}
\\ 
\hline
\multicolumn{1}{r|}{$\psi_0(r)$\footnotemark[1]} & 
2.238(4) & 15.9(2) & -0.3491(4) & 0.1122(4)  & -0.6067(12) \\
\multicolumn{1}{r|}{$\overline{\psi}_1(\mathbf{r})$\footnotemark[2]} & 
0.19775(14) & 6.060(10) & 0.16535(9) & 1.907(2) & \\
\hline
\hline
\multicolumn{1}{l}{$Mg$} & & & & & \\
\hline
\hline
geminal &
\makebox[0pt][c]{$Z_1$} & \makebox[0pt][c]{$Z_2$}
& \makebox[0pt][c]{$p$}   & \multicolumn{1}{c}{$\lambda$} & \\ 
\hline
\multicolumn{1}{r|}{$1s$} & 12.0    & 0.0   &   \mbox{\#} 
& ~1.0 & \\
\multicolumn{1}{r|}{$2s$} & 10.784(2)   & 3.1685(4)  &   \mbox{\#} 
& 2.40(18)~\mbox{$10^{-1}$} & \\
\multicolumn{1}{r|}{$2p$} & 6.6199(4)   & 3.25296(19)  & ~0.19285(9) 
& 8.56(2)~\mbox{$10^{-1}$} & \\
\multicolumn{1}{r|}{$3s$}  &  1.9220(3)  & 1.01249(15) & 0.08312(8) 
& 1.674(13)~\mbox{$10^{-4}$} & \\
\multicolumn{1}{r|}{$3p$}  & 2.599(12)  & 1.2738(3) & 0.535(14) 
&  -2.152(17)~\mbox{$10^{-5}$} &  \\ 
\hline
\hline
Jastrow 2 body & \multicolumn{1}{c}{$b$} 
& \multicolumn{4}{c}{} \\
\hline
  &  1.191(5) & & & & \\
\hline
\hline
Jastrow 3 body &  
 \makebox[0pt][c]{$Z_1$} & \makebox[0pt][c]{$Z_2$}
& \makebox[0pt][c]{$a_0$}   & \multicolumn{1}{c}{$a_1$} 
& \multicolumn{1}{c}{$a_2$} \\
\hline
\multicolumn{1}{r|}{$\psi_0(r)$\footnotemark[3]} & 
1.465(6) &   & -0.3919(5) & 1.3867(9) & -0.891(2) \\
\multicolumn{1}{r|}{$\overline{\psi}_1(\mathbf{r})$\footnotemark[2]} & 
13.03(3) & 1.074(10) & 0.8789(2) & 0.29822(5) & \\
\hline
\hline
\end{tabular} 
\footnotetext[1]{$\psi_0(r)=a_0 ( \exp(-Z_1 r^2) + a_1 \exp(-Z_2r^2) + a_2 ) $}
\footnotetext[2]{$\overline{\psi}_1(\mathbf{r})=\mathbf{r}~ a_0 ( \exp(-Z_1 r^2) + 
a_1 \exp(-Z_2 r^2) )$}
\footnotetext[3]{$\psi_0(r)=a_0 ( (1 + a_1 r^2)  \exp(-Z_1 r^2) + a_2)$}

%\end{ruledtabular}
}
\end{table*}

\end{document}